\documentclass[%
 aps
 amsmath,amssymb,
 reprint,
]{revtex4-1}

\usepackage{graphicx}%
\usepackage{dcolumn}%
\usepackage{bm}%

\usepackage[utf8]{inputenc}
\usepackage[T1]{fontenc}
\usepackage{mathptmx}
\usepackage{siunitx}
\usepackage{hyperref}
\newcommand{\gammam}{\gamma_\text{m}}

\begin{document}

\preprint{APS/123-QED}

\title[Gigahertz phononic integrated circuit]{Gigahertz phononic integrated circuits on thin-film lithium niobate on sapphire}

\author{Felix M. Mayor}
\thanks{These authors contributed equally to this work}
\author{Wentao Jiang}%
\thanks{These authors contributed equally to this work}
\author{Christopher J. Sarabalis}
\author{Timothy P. McKenna}
\author{Jeremy D. Witmer}
\author{Amir H. Safavi-Naeini}
 \email{safavi@stanford.edu.}
\affiliation{ 
Department of Applied Physics and Ginzton Laboratory, Stanford University, 348 Via Pueblo Mall, Stanford, California 94305, USA%
}%

\date{\today}%

\begin{abstract}
Acoustic devices play an important role in classical information processing. The slower speed and lower losses of mechanical waves enable compact and efficient elements for delaying, filtering, and storing of electric signals at radio and microwave frequencies. Discovering ways of better controlling the propagation of phonons on a chip is an important step towards enabling larger scale phononic circuits and systems. We present a platform, inspired by decades of advances in integrated \emph{photonics}, that utilizes the strong piezoelectric effect in a thin film of lithium niobate on sapphire to excite guided acoustic waves immune from leakage into the bulk due to the phononic analogue of index-guiding. We demonstrate an efficient transducer matched to $\SI{50}{\ohm} $ and guiding within a $\SI{1}{\micro\meter}$ wide mechanical waveguide as key building blocks of this platform. Putting these components together, we realize  acoustic delay lines, racetrack resonators, and meander line waveguides for sensing applications. To evaluate the promise of this platform for emerging quantum technologies, we characterize losses at low temperature and measure quality factors on the order of $50,000$ at $\SI{4}{\kelvin}$. Finally, we demonstrate phononic four-wave mixing in these  circuits and measure the nonlinear coefficients to provide estimates of the power needed for relevant parametric processes.

\end{abstract}

\maketitle

\section{Introduction}
\label{sec:intro}

Photonic integrated circuits~\cite{thomson2016roadmap,rahim2017expanding,marpaung2019integrated} have allowed integration of on-chip optical components to realize optical systems with unprecedented performance. This has largely been enabled by high-confinement waveguides that are compact, allow for tight bends, and localize energy into a small mode area for efficient interactions. However, similar capabilities do not yet exist for phononic circuits (hereafter, phononic is used interchangeably with the words acoustic and mechanical). The ability to guide, control, and mix acoustic waves may enable important new devices to support development of classical and quantum information systems. Phononic circuits have been developed on a number of candidate platforms. Following the much earlier pioneering work on surface acoustic wave devices~\cite{oliner1976waveguides,weller1977surface,Auld1990v2,campbell2012surface}, the last decade has brought about a number of new platforms including suspended~\cite{shin2013tailorable,van2018electrical,liu2019electromechanical,sarabalis2020s,dahmani2020piezoelectric} or unreleased~\cite{pant2011chip,van2015interaction,sarabalis2017release,liu2017toward,fu2019phononic} rib waveguides, as well as membrane~\cite{hatanaka2014phonon,romero2019propagation} and phononic crystal~\cite{khelif2006complete,vasseur2007waveguiding,olsson2008microfabricated,mohammadi2009high,khelif2010acoustic,fang2016optical,balram2016coherent,patel2018single} based waveguides. 

To further realize the potential of phononic integrated circuits, we seek a platform that combines broadband and efficient electromechanical wave transduction with a robust and scalable guiding mechanism that allows for wavelength-scale localization of acoustic energy. Index guiding, the predominant guiding mechanism in \emph{photonic} integrated circuits, relies on slower wave speed in the guiding material than the surroundings. The same mechanism can also be realized in phononic circuits~\cite{safavi2019controlling,liu2017toward,sarabalis2016guided,poulton2013acoustic,Wang2020} with careful selection of the materials comprising the platform. We demonstrate in this work a new platform for phononic circuits, LiNbO$_3$-on-sapphire (LiSa), that combines the high piezoelectric coupling coefficient of lithium niobate with the fast wave propagation in sapphire to realize both of the aforementioned requirements. The fast mechanical wave speed in sapphire ($\sim\SI{6.4}{\kilo\meter\per\second}$) enables guiding in a wavelength-scale rib waveguide in lithium niobate (LN), where the guided mechanical wave propagates at $ \sim\SI{3.3}{\kilo\meter\per\second} $. The unreleased nature of phononic circuits on LiSa eases fabrication of large-scale dense circuits, while the evanescent mode profile in sapphire provides controllable coupling between adjacent waveguides.
Moreover, the higher refractive index of LN ($\sim 2.2$) compared to sapphire ($\sim 1.7$) allows for simultaneous optical guiding in the same waveguide~\cite{mckenna2020cryogenic}. Implementing phononic circuits in a material platform that also supports photonics opens up ways for more complex electro- and piezo-optomechanical circuits~\cite{safavi2019controlling,sarabalis2020acousto}. %
Here, we demonstrate gigahertz-frequency phononic integrated circuits with wavelength-scale confinement and efficient piezoelectric transduction on an unreleased platform. We present the design of an interdigital transducer (IDT) on LiSa in Sec.~\ref{sec:design} and demonstrate via simulations the mode structure of the waveguide and the electromechanical response of the transducer. The high dielectric constant of LN ($\epsilon_r\sim 30$) and the high piezoelectric coupling ($k^2_\text{eff}\sim15\%$) of the transducer mode enable a compact transducer with an active area of $ \sim\SI{2}{\micro\meter}\times\SI{10}{\micro\meter}$. In Sec.~\ref{sec:char}, we present measurements of the performance of the transducer and extract the properties of the phononic waveguide. We utilize the transducer and the waveguide to compose components of phononic integrated circuits, including racetrack resonators (Sec.~\ref{sec:racetrack}) and meander waveguides to demonstrate two-dimensional (2D) reflectometry (Sec.~\ref{sec:2D-refl}). To better understand the loss mechanisms affecting these circuits, we measure phononic racetrack resonators at both room temperature and at $\sim \SI{4}{\kelvin}$. Lastly, in Sec.~\ref{sec:FWM}, we demonstrate the four-wave mixing nonlinearity in the mechanical waveguide and extract the modal nonlinear coefficient, from which we estimate a sub-milliwatt threshold pump power for observing parametric amplification. Potential applications of the gigahertz phononic integrated circuits are discussed in section~\ref{sec:conclusion}.

\section{Design}
\label{sec:design}

\begin{figure*}[ht]
  \includegraphics[scale=1]{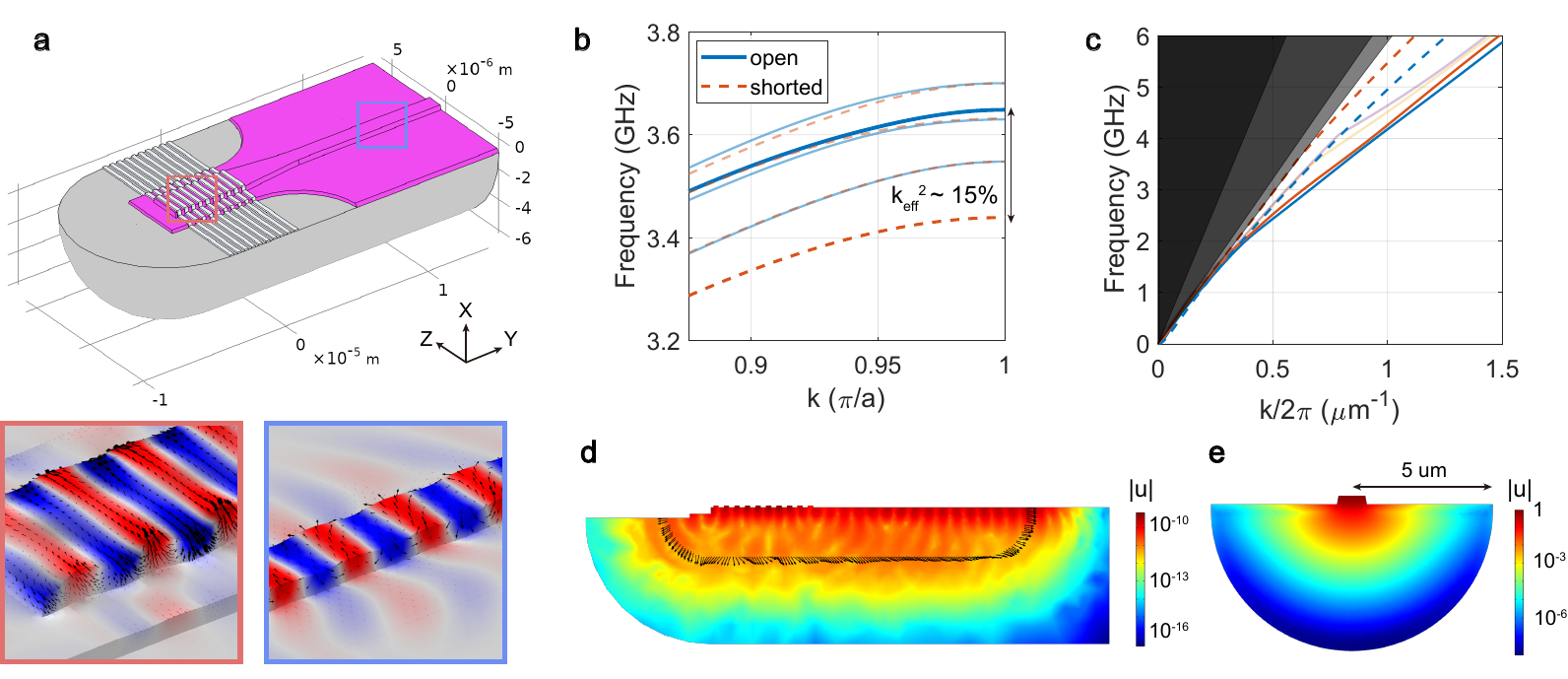}
\caption{\label{Fig1:design} Design and simulation of the IDT and the mechanical waveguide. (a) Finite-element model and simulated driven mode profile of the IDT-taper-waveguide system. The displacement is shown as black arrows, and the color highlights the transverse horizontal component of the displacement field. (b) First four bands of the IDT unitcell with open (solid blue) and shorted  (dashed red) electrode boundary condition. The strongly coupled mode is emphasized. (c) Mechanical bands of the LN waveguide (solid) and the LN slab (dashed). Black cones represent bulk acoustic and surface acoustic waves in sapphire. (d) Displacement field on a two-dimensional cross section of the IDT-taper-waveguide system. Black arrows indicate the direction of mechanical energy flux. (e) Normalized displacement field of the quasi-Love guided mode in the rib waveguide.}
\end{figure*}

An efficient microwave-to-mechanical transducer needs to have an input impedance close to $\SI{50}~\Omega$ over its bandwidth $\gamma$ to absorb most of the incoming electrical power. It should also convert this absorbed power into mechanical waves in a desired mode with high efficiency. The impedance matching condition can be satisfied by increasing the area of the transducer~\cite{sarabalis2020s}. But larger transducers have modes that are not as well matched to small phononic waveguides. To efficiently excite a wavelength-scale waveguide, previous demonstrations have either shaped the  transducer to focus  the emission~\cite{siddiqui2018lamb,liu2019electromechanical} or tapered the region between the waveguide and transducer~\cite{dahmani2020piezoelectric}. The latter scheme, which we pursue  here, works best with transducers that are not significantly wider than the waveguide they are exciting. Large electromechanical coupling enables the efficient and small transducers needed to realize these devices. 

The high piezoelectric coefficient and dielectric constant of LN enable the impedance matching with a small total transducer area. We choose the IDT width to be $ w_\text{IDT} = \SI{2}{\micro\meter}$, close to the waveguide width of $\SI{1}{\micro\meter}$, so that it can be connected to the waveguide with a simple linear taper~\cite{dahmani2020piezoelectric}. To utilize the largest piezoelectric coefficient of LN ($d_{24} = d_{15} \sim \SI{70}{\pico\coulomb\per\newton} $~\cite{weis1985lithium}), we look for IDT configurations that couple the $XY$ strain component to the electric field along $Y$ direction, where $XYZ$ denote the crystal axes. Such a configuration can be realized on $X$-cut lithium niobate by orienting the direction of propagation to be parallel to the crystal $Y$ axis, as shown in Fig.~\ref{Fig1:design}a, where the shear horizontal (SH) mode in the IDT region is strongly transduced via the $d_{24}$ component. The LN (purple) slab layer surrounding the IDT region is removed to suppress the excitation of surface and bulk acoustic waves.

We use finite element method~\cite{COMSOL} to simulate the IDT unitcell with periodic boundary condition to estimate the transducer parameters, including the frequency of the strongly coupled SH mode, the piezoelectric coupling coefficient $k^2_\text{eff}$, and the required total area for impedance matching. We choose an IDT periodicity $a_\text{IDT} = \SI{1}{\micro\meter}$ (setting the mechanical wavelength to \SI{1}{\micro\meter}), and an electrode duty cycle of $50\%$, which corresponds to electrode width and gap of both $\SI{250}{\nano\meter}$. To estimate the coupling coefficient $k^2_\text{eff}$, we simulate how the mode frequencies shift when the boundary condition on the electrodes is changed from open to short. Fig.~\ref{Fig1:design}b shows the first four bands of the IDT unitcell near the $X$-point with open (solid blue) and shorted (dashed red) boundary condition on the electrodes. Modes with weak piezoelectric coupling show a weak dependence on the electrode boundary condition, while the mode with strong coupling shows a large frequency shift (highlighted) when the boundary condition is changed. The fractional difference in frequency is approximately given by $k^2_\text{eff}$~\cite{dahmani2020piezoelectric}. We identify the SH mode of the IDT from both the mode profile and its large piezoelectric $k_\text{eff}^2 \approx 15\%$. The dielectric capacitance $C_0$ of the unitcell is simulated by solving the model at low frequencies ($\sim\SI{}{\kilo\hertz}$) where its admittance $Y(\omega) \approx i\omega C_0$. We obtain $C_0 = \SI{5.4}{\femto\farad}$ per unitcell. The required area of the IDT for reaching a certain conductance $G_0$ over a bandwidth $\gamma$ can be estimated by the formula~\cite{sarabalis2020s,dahmani2020piezoelectric}
\begin{equation}
    A = \frac{\pi^2}{8}\frac{G_0}{\omega^2 c_0}\frac{\gamma}{k_\text{eff}^2},
\end{equation}
where $c_0$ is the capacitance per unit area and $G_0$ is the peak conductance. Aiming to match over a broad bandwidth to a $\SI{50}{\ohm}$ transmission line, we set $G_0 = \SI{20}{\milli\siemens}$ and $\gamma/2\pi \sim \SI{50}{\mega\hertz}$. The bandwidth of $\gamma/2\pi \sim \SI{50}{\mega\hertz}$ is chosen to be large enough to exceed the intrinsic dissipation of LN IDTs at room temperature (measured $Q\sim 400$, implying $\approx\SI{10}{\mega\hertz}$ loss-limited bandwidth), while small enough to avoid requiring techniques such as chirping which are needed for phase-matching over very large bandwidths~\cite{manzaneque2017lithium}.
This leads us to estimate the required number of unitcells to be $N\sim 10$ corresponding to a transducer area of $A\sim \SI{20}{\micro\meter\squared} $. From simulations of the full IDT-taper-waveguide system, we are able to identify the SH motion in both the IDT region and in the waveguide, verifying our design. Insets in Fig.~\ref{Fig1:design}a show the mode profiles in the IDT and waveguide regions when the IDT is driven at its resonant frequency. The simulated admittance versus frequency is presented in Appendix~\ref{app:IDT_admittance} and shows fairly good agreement with measurements presented later in the text.

To identify the guided acoustic waves in the rib waveguide, we perform a quasi-two-dimensional simulation of the waveguide cross section. Fig.~\ref{Fig1:design}c shows the band diagram for mechanical waves in the LiSa rib waveguide. A waveguide thickness $t = \SI{300}{\nano\meter}$ and width $ w = \SI{1}{\micro\meter} $ is chosen with a slab layer of thickness $ t_\text s = \SI{200}{\nano\meter} $. The first four bands of the waveguide are plotted as solid lines. The $\SI{200}{\nano\meter}$ LN slab on sapphire also supports Rayleigh and Love waves that are shown as dashed lines. The longitudinal and shear waves in bulk sapphire, as well as the surface acoustic wave (SAW) on the sapphire-air interface are shown as the grey shaded regions. We observe that at frequencies above $\sim \SI{2}{\giga\hertz}$, the first two bands of the guided waves, which correspond to the quasi-Love and the quasi-Rayleigh modes have similar group velocities and are well separated from the bulk- and slab-modes. The coincidental proximity of the two bands causes them to hybridize. This hybridization can be avoided for other waveguide orientations or geometries where the two bands are well separated. To illustrate the tight confinement of the guided mechanical mode, we show the normalized displacement $|u|$ of the quasi-Love mode at $3.5$ GHz on the waveguide cross section in Fig.~\ref{Fig1:design}e. The displacement drops by five orders of magnitude $\SI{5}{\micro\meter}$ away from the center of the waveguide.

In contrast to a released IDT where material loss dominates the efficiency at room temperature~\cite{dahmani2020piezoelectric}, the unreleased IDT may also emit energy into the substrate as bulk acoustic waves (BAW, see Appendix~\ref{app:BAW}). These losses would limit the conversion efficiency of the transducer, even in the absence of material loss. We show the simulated displacement field on a cross section of the IDT-taper-waveguide system along the direction of propagation in Fig.~\ref{Fig1:design}d, showing emission of acoustic waves into the sapphire. From simulations, we infer that out of all the mechanical power generated by the transducer, $64\%$ is guided in the modes of the rib waveguide, $14\% $ is lost in the slab layer, and the rest is radiated into the sapphire substrate. This means that the maximum efficiency of the IDT is $\sim 64\% $. The actual efficiency may additionally suffer from impedance mismatch between the IDT and the microwave transmission line, resistive loss of the metal electrodes, and material loss of LN.

\section{Transducer Fabrication and characterization}
\label{sec:char}

\begin{figure}[htbp]
  \includegraphics[scale=1]{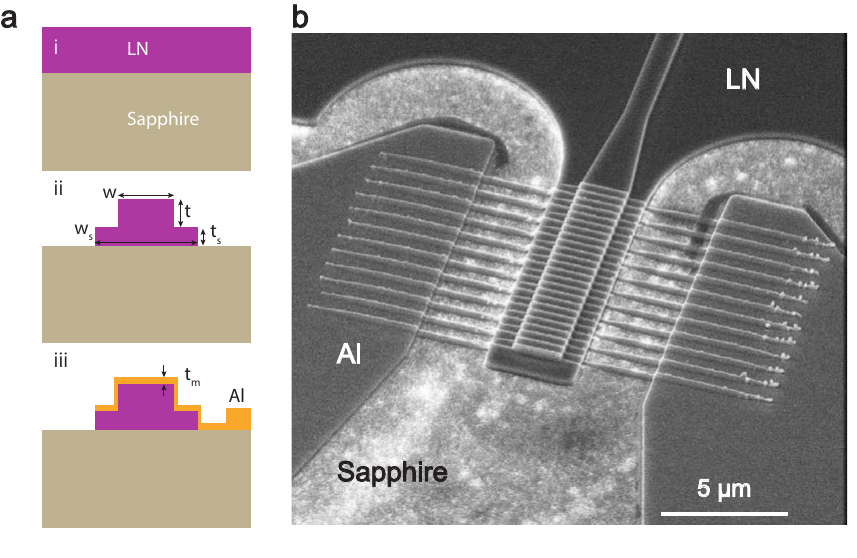} 
\caption{\label{Fig2:fab} (a) Fabrication process of the LiSa IDT. (b) Scanning electron micrograph of one fabricated transducer. }
\end{figure}

The fabrication process is illustrated in Fig.~\ref{Fig2:fab}a. We start with $\sim \SI{500}{\nano\meter}$ thin-film LN on sapphire. A two-step etching process is adopted to define the IDT structure. The IDT metallization layer is patterned by electron beam lithography and liftoff. A detailed description of the fabrication process is presented in Appendix~\ref{sec:fab}. Fig.~\ref{Fig2:fab}b shows a scanning electron micrograph (SEM) of the fabricated transducer. The etching and cleaning steps likely cause roughness on the sapphire surface, which is accentuated by charging during the SEM. This roughness should have little effect on device performance since the mechanical waves do not interact with the bare sapphire surface.

When we apply a microwave drive to an IDT, part of the signal is reflected due to impedance mismatch between the IDT and the $\SI{50}{\ohm}$ transmission line. Near the resonance of the IDT, the microwave reflection $S_{11}$ drops where microwave energy is converted to mechanical energy. As explained in the previous section, only part of this energy is converted to propagating phonons in the waveguide. We characterize the efficiency of the IDT using the scattering parameter $t_{\text m \mu}$ between the microwave input amplitude and the propagating mechanical wave amplitude in the rib waveguide. The propagating phonons reach a second IDT, where they are either reflected, scattered into BAW, dissipated by material loss, or converted back to microwaves. This allows us to measure the microwave transmission $S_{21}$ through the IDT-waveguide-IDT system. Reflection of the phonons at the IDT-waveguide interface causes the appearance of fringes on the microwave reflection and transmission spectra. These periodic fringes in  the spectrum correspond to echoes in the time-domain impulse response.

\begin{figure}[htbp]
  \includegraphics[scale=0.99]{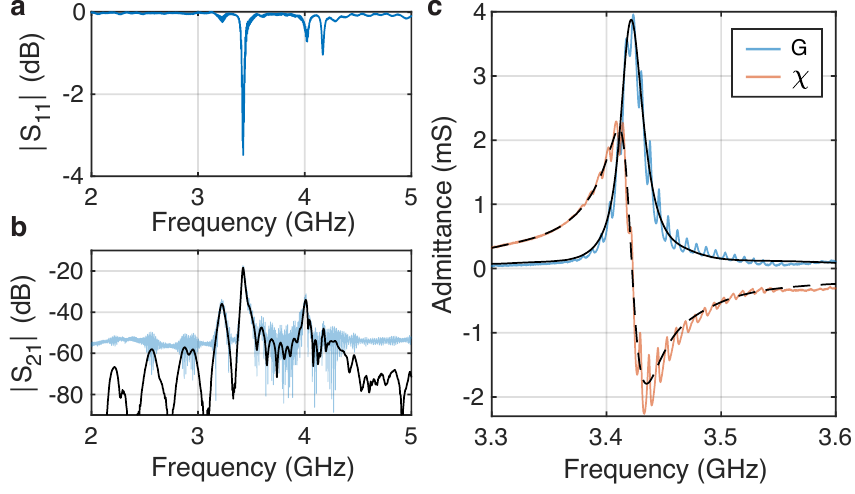} 
\caption{\label{Fig:IDT-measurement} (a) Microwave reflection $\left|S_{11}\right|$ of a transducer with $N=9$ unitcells. (b) Transmission $\left|S_{21}\right|$ of an IDT-waveguide-IDT system where the waveguide length is $L=\SI{200}{\micro\meter}$. The blue curve is the unfiltered response and the black curve has been filtered to keep only the single transit response. (c) Measured conductance $G$ (blue) and susceptance $\chi$ (red) of the transducer. The filtered response is shown in black and corresponds to the transducer response without mechanical reflections from the other transducer.}
\end{figure}

The microwave scattering parameters $S_{ij}$  of the whole device are measured with a vector network analyzer (VNA). The VNA (Rhode \& Schwartz, ZNB$20$) is calibrated up to the tips of electrical probes (GGB Industries, model $40$A) that are used to apply the microwave drive on the transducers. Here, we present typical results for a device with a waveguide length $L=\SI{200}{\micro\meter}$ and $N=9$ unitcells for each transducer, measured at room temperature. In reflection, we see that the main lobe of the IDT is centered around $\omega/2\pi = 3.42$ GHz with $\left|S_{11}\right|^2=0.449~(\SI{-3.48}{\deci\bel})$ (see Fig.~\ref{Fig:IDT-measurement}a). In Fig.~\ref{Fig:IDT-measurement}c, we show the measured admittance associated with this IDT lobe. A Lorentzian fit allows us to extract a bandwidth of $\gamma/2\pi = 23$ MHz and a peak conductance $G_0 = \SI{3.9}{\milli\siemens}$.

We measure the delay line in transmission (Fig.~\ref{Fig:IDT-measurement}b) and find a maximal $\left|S_{21}\right|^2 = \SI{0.018}{}~(\SI{-17.4}{\deci\bel})$. Naively, one would expect that the total fraction of the input power transmitted through the device, $\left|S_{21}\right|^2$, is simply the product of the two IDTs' conversion efficiency $T=|t_{\text{m}\mu}|^2$, so $|S_{21}|^2=T^2$.  However, this approximation is not accurate, even if the two IDTs were identical. Two additional factors must be considered: the propagation loss $\alpha$ along the  waveguide, and the mechanical reflections $R$ at the waveguide-transducer interface. The latter may resonantly enhance transmission through the waveguide.  

To independently determine reflection ($R$) and conversion efficiency ($T$) of the IDT as well as waveguide propagation loss, we fabricate and perform measurements on devices with five different waveguide lengths $L=0.4$, $0.6$, $0.8$, $1.1$ and $\SI{1.5}{\milli\meter}$. In the time-domain impulse response $h_{21}(t)$ of every  device, we observe peaks ($h[0], h[1], h[2],\dots, h[n],\dots$) at different delays, indexed according to the number of successive echoes $n$ matching that delay. We restrict ourselves to $n\leq2$ where the echoes can be clearly identified. Using the data from devices with varying $L$, and the dependence of the peak heights on $R, T, \alpha$, and  $n$ as explained in Appendix~\ref{sec:deembed},  we obtain the transducer and waveguide parameters $R=32 \pm 8$\%, $T \approx 13$ \%, and $\alpha = 4.0 \pm \SI{0.2}{\deci\bel\per\milli\meter}$. This propagation loss is comparable to previous measurements of LN SAW resonators at similar frequencies~\cite{shao2019phononic}. Additionally, while for an ideal IDT we expect $R+T=1$, losses in the transducer lead to $R+T<1$.

We measure a delay of $\SI{453 \pm 3}{\nano\second}$ in a waveguide with length $L=\SI{1.5}{mm}$. This  corresponds to a group velocity of $v_\text{g,meas} = \SI{3320 \pm 30}{\meter\per\second}$. From simulations, we expect $v_\text{g,R} = \SI{3336}{\meter\per\second}$ for the quasi-Rayleigh mode and $v_\text{g,L} = \SI{3541}{\meter\per\second}$ for the quasi-Love mode. Fabrication uncertainties on waveguide thickness and slab thickness, and more importantly the IDT and waveguide orientation, could significantly affect the wave speeds. To provide further evidence that the mechanical waves are guided, we fabricated several experimental control devices where the two IDTs are not interconnected by a mechanical waveguide, and measured a peak $ S_{21}$ below $\SI{-50}{\deci\bel}$. %

\section{Acoustic racetrack resonators}
\label{sec:racetrack}
In photonic integrated circuits, ring (or racetrack) resonators~\cite{bogaerts2012silicon} are widely used for filtering and sensing. A ring resonator consists of a waveguide that is fed back to itself. This leads to the formation of resonances separated in frequency by a free spectral range (FSR) of $\Delta f = v_g/L$, where $v_g$ is the group velocity in the waveguide and $L$ is the round-trip length of the ring resonator. Here, we implement the phononic analogue of a racetrack resonator. These resonators may be used as microwave filters with narrow linewidths. At cryogenic temperatures, high quality factors and robustness to environmental factors (such as optical photons, high-energy particles, and magnetic fields) may make them attractive alternatives to superconducting microwave resonators which may be prone to quasiparticle induced losses~\cite{barends2011minimizing}. Due to the shorter wavelength of the mechanical waves compared to microwaves, we also have access to a larger number of modes, given limited space. Multiple resonators can be cascaded to achieve larger control over the spectrum of the filter, for example to obtain high order band-pass filters~\cite{cleland2019mechanical}.

\begin{figure*}[htbp]
  \includegraphics[scale=1]{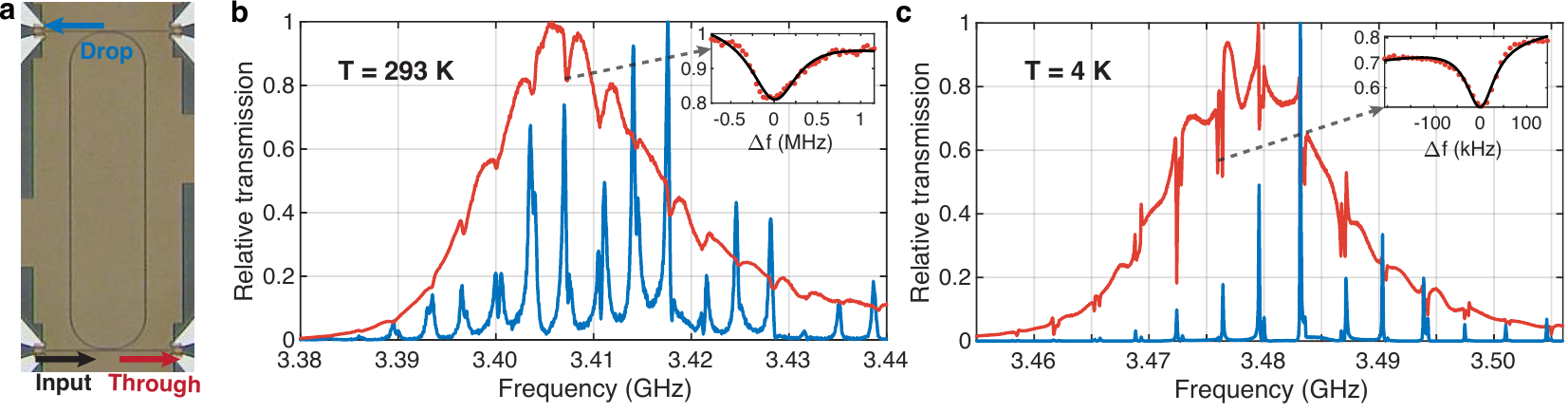} 
\caption{\label{Fig:racetrack} Acoustic racetrack resonator. (a) Optical microscope image of an acoustic  racetrack resonator. The input as well as the through and drop ports are labeled. (b), (c) Relative power transmission at room temperature (b) and $T\approx 4$~K (c) of the through (red) and drop (blue) channels. The insets show typical acoustic resonances with quality factor $Q = 5,000$ at room temperature~(b) and $Q = 47,000$ at $4$~K~(c).}
\end{figure*}

Coupling to the racetrack resonator is achieved by evanescent coupling to a nearby waveguide. Because the displacement of the confined mode decays exponentially inside the slab and sapphire (Fig.~\ref{Fig1:design}e), the coupling also displays an exponential dependence on the waveguide-to-resonator separation gap $\tilde{g}$, allowing us to engineer the coupling rate via both $ \tilde{g} $ and the length of the coupling section $L_\text c$. A fabricated racetrack resonator in an add-drop filter configuration  is shown in Fig.~\ref{Fig:racetrack}a with IDTs at each of the four ports. The racetrack has a total length $L=\SI{871}{\micro\meter}$ with bending radius $R_\text{b}=\SI{40}{\micro\meter}$. The coupling between the waveguide and the resonator is identical on both sides, with a coupling length $L_\text{c}=\SI{10}{\micro\meter}$ and gap $\tilde{g}=$ \SI{800}{nm}.

We measure the scattering parameters of the racetrack resonator inside a closed-cycle cryostat (Montana Instruments) at both room temperature (Fig.~\ref{Fig:racetrack}b) and $T\approx \SI{4}{\kelvin}$ (Fig.~\ref{Fig:racetrack}c) in the through and drop configurations (labelled in Fig.~\ref{Fig:racetrack}a). The scattering parameters are filtered to reduce crosstalk (see Appendix~\ref{app:filter}). In the through (drop) configuration, the dips (peaks) correspond to resonances of the racetrack. These are separated by the FSR $\Delta f = \SI{3.51 \pm 0.06}{\mega\hertz}$ at room temperature and $\Delta f = \SI{3.57 \pm 0.02}{\mega\hertz}$ at $T=\SI{4}{\kelvin}$. From the FSR, we infer a mean group velocity inside the racetrack resonator of $v_\text{g,meas} = \SI{3.1}{\kilo\meter\per\second}$. This is lower than the measured $v_\text g$ of the straight waveguide considered in the previous section. The discrepancy between the two values is due to the highly anisotropic nature of LN and the dependence of the group velocity on crystal orientation. We take this dependence into account and calculate the average group velocity seen by waves propagating along the racetrack to find $v_\text{g,R}=\SI{3.1}{\kilo\meter\per\second}$ and $v_\text{g,L}=\SI{3.0}{\kilo\meter\per\second}$ for the quasi-Rayleigh and quasi-Love modes. These lower values are in qualitative agreement with the measurement $v_\text{g,meas}$. In addition, we observe backscattering induced splitting of the resonances~\cite{li2016backscattering} on the order of $400$ kHz, limiting our ability to precisely determine  the  FSR. Finally, we note that in addition to the temperature dependence of the FSR, the transducer center frequency shifts up by $\sim 70$ MHz when going from room temperature to $T=\SI{4}{\kelvin}$.

The resonances of the racetrack allow us to extract the quality factors and infer the corresponding propagation losses. We obtain intrinsic quality factors $Q_\text{i} = 6,500 \pm 1,700$ at room temperature and $Q_\text{i} = 46,000 \pm 9,000$ at $T=\SI{4}{\kelvin}$. These correspond to propagation losses $\alpha = 4.9 \pm \SI{1.6}{\deci\bel\per\milli\meter}$ and $\alpha = 0.7 \pm \SI{0.2}{\deci\bel\per\milli\meter}$ where $\alpha = \omega/(Q_\text{i} v_\text{g})$. The propagation loss at room temperature is consistent with the one measured from the time-domain impulse response in section \ref{sec:char}. At $T=\SI{4}{\kelvin}$, the propagation loss goes down by an order of magnitude. This suggests that thermally induced losses limit the quality factors at room temperature.
We also infer from simulations detailed in Appendix~\ref{sec:bending} that the quality factor of the racetrack resonator is not limited by bending loss. Though we are not currently certain of the source of loss at 4 kelvin, we expect that with better fabrication and materials processing, and by going to lower temperatures, significantly higher mechanical quality factors and correspondingly smaller attenuation constants can be attained.

\section{Meander waveguide Two-dimensional Reflectometry}
\label{sec:2D-refl}

Acoustic microscopy has long been used for non-destructive imaging~\cite{jipson1978acoustic} and crack detection~\cite{kino1978application,khuri1980acoustic}. More recently, remarkable progress in integrated nanoelectromechanical (NEM) resonators has enabled mass spectrometry with unprecedented sensitivity~\cite{ekinci2004ultimate} and scale~\cite{sage2018single}, enabled by the low effective mass of these nanoresonators and their integration on a chip. The phononic waveguides shown here possess the high sensitivity and integration potential of NEM systems, while also being suitable for schemes  based on time-domain reflectometry that have previously been in the domain of SAW and BAW devices.

\begin{figure}[htbp]
  \includegraphics[scale=0.89]{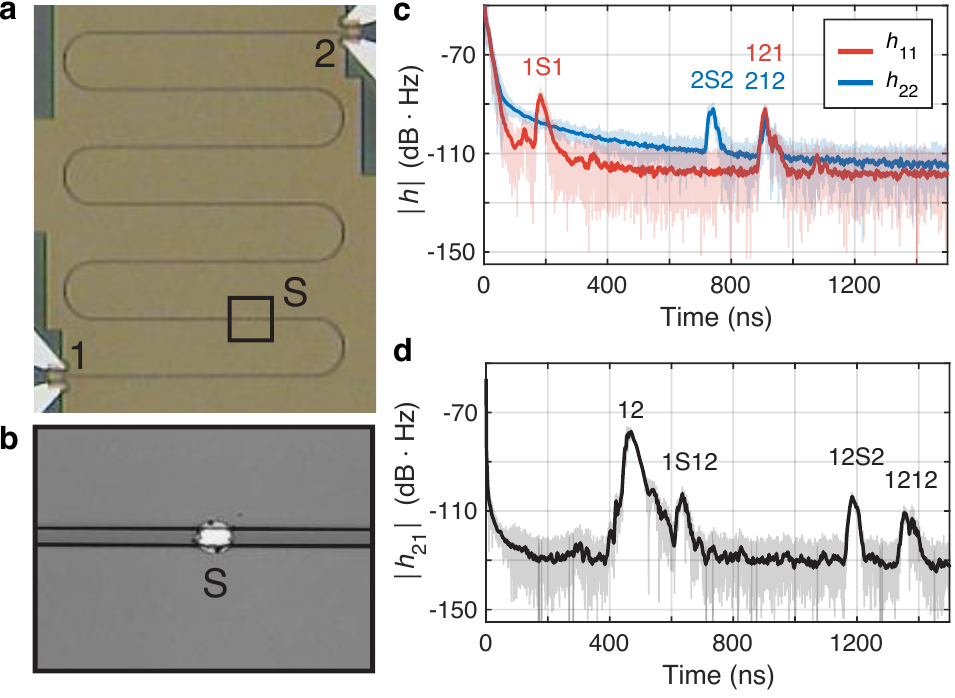} 
\caption{\label{Fig:refl} (a) Optical microscope image of a meander waveguide used for two-dimensional reflectometry experiments. The input and output ports are labeled as $1$ and $2$. The aluminum scatterer on the waveguide is labeled as S. (b) Optical microscope image showing a close-up of the scatterer. (c) Impulse response $h(\tau)$ in reflection for both transducers. The peaks are labeled with the corresponding sequence of scattering events. The highlighted curves have been smoothed with a $50$-point (\SI{5}{\nano\second}) moving average. (d) Impulse response $h(\tau)$ in transmission.}
\end{figure}

By sending out a signal and measuring the time of return and amplitude of the reflected signal, we can estimate the position and mass of a scatterer. Approaches to obtaining a two-dimensional map of a surface include scanning an acoustic probe across a surface (scanning acoustic microscopy), using multiple SAW transducers patterned on the surface to perform triangulation-based time-domain reflectometry~\cite{lee2005apparatus}, or using multiple modes of a single resonator~\cite{hanay2012single}. Here, we utilize the guided nature of the mechanical wave to cover a two-dimensional region with a single meander line waveguide. The high confinement provided by the waveguide allows enhanced mass sensitivity as well as the tight bends needed to implement the meander line.

Our meander waveguide sensor has a total length $L=1.5$ mm and a bending radius $R_\text{b}=$\SI{20}{\micro\meter} (see Fig.~\ref{Fig:refl}a). To test the sensor, we deposit an aluminum scatterer (Fig.~\ref{Fig:refl}b) of radius $R_\text{s}=\SI{1}{\micro\meter}$ and thickness $t_\text{m}=\SI{100}{\nano\meter}$ on the waveguide at a propagation distance $\Delta z_\text{s}=\SI{286}{\micro\meter}$ away from IDT $1$.  For each path that the acoustic wave can take in the meander line, we observe a peak in the impulse response with the corresponding delay that is proportional to the total propagation distance. Among these paths, the ones where one reflection at the scatterer is involved allow us to determine the position of the scatterer.

We compute the time-domain impulse response $\left|h_{ij}(\tau)\right|$ of the device from the scattering parameters $S_{ij}(\omega)$ measured with a VNA in frequency-domain. In Fig.~\ref{Fig:refl}c we show the one-port reflection impulse response $\left|h_{11}\right|$ and $\left|h_{22}\right|$ for IDT~$1$ and $2$ respectively, where we label the peaks with the corresponding sequence of scattering events. The delay $\tau_\text{121} = \tau_\text{212} = \SI{909}{\nano\second}$  is identical for both IDTs and does not involve the scatterer, while the other delay is clearly separated and set by the location of the scatterer relative to the IDT. From the delays $\tau_\text{1S1} = \SI{182}{\nano\second}$ and $\tau_\text{2S2} = \SI{736}{\nano\second}$, we compute the expected position of the scatterer to be $\Delta z_\text{s,meas1} = L \tau_\text{1S1}/\tau_\text{121} =\SI{302}{\micro\meter}$ and $\Delta z_\text{s,meas2} = L \left(1 - \tau_\text{2S2}/\tau_\text{212}\right)= \SI{288}{\micro\meter}$. It is also possible to determine the position of the scatterer from the transmission impulse response $h_{21}$ (see Fig.~\ref{Fig:refl}d). We observe four peaks, two of which correspond to a path involving the scatterer. We find $\Delta z_\text{s,meas3} = \SI{275}{\micro\meter}$ from the first peak, and $\Delta z_\text{s,meas4} = \SI{281}{\micro\meter}$ from the second. Averaging over the measured values, we obtain $\Delta z_\text{s,meas} = \SI{286\pm 12}{\micro\meter}$ which matches the actual position $\Delta z_\text{s}$. We are thus able to confirm the location of the scatterer in the two-dimensional region spanned by the meander line. We note that the spacing between the straight waveguide sections are currently set by twice the bending radius, which can be reduced to $ \sim \SI{5}{\micro\meter} $ with a more careful design, while still avoiding inter-waveguide coupling. The resolution between different scattering events in time-domain could also be improved by increasing the transducer bandwidth. A study of the limits of mass sensitivity for such  a structure  will be performed in future work.

\section{Low power acoustic four-wave mixing}
\label{sec:FWM}

The small effective mode area $\sim \SI{0.1}{\micro\meter^2}$ of our waveguide allows the observation of nonlinear mechanical effects with comparatively lower powers. In the past few decades, there have been numerous demonstrations of nonlinear interactions between mechanical waves in bulk acoustic waves~\cite{thompson1970acoustic,thompson1971nonlinear}, surface acoustic waves~\cite{naianov1986surface} and more recently in phononic waveguides and resonators~\cite{kurosu2018chip,kurosu2020mechanical,mahboob2011wide}. Here, we focus on four-wave mixing (FWM) where two phonons are annihilated to create two phonons.

\begin{figure}[htbp]
  \includegraphics[scale=1]{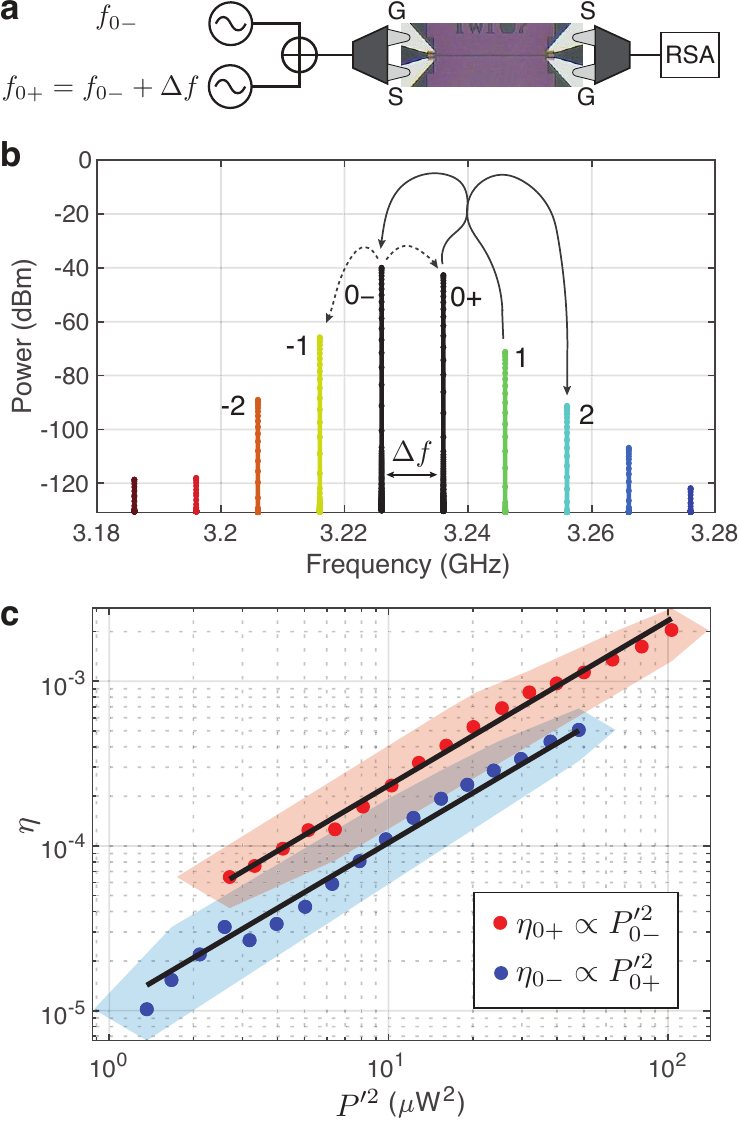} 
\caption{\label{Fig:Nonlinear} (a) Four-wave mixing measurement setup. (b) Measured power for each comb line with $\Delta f = 10$ MHz. The two central black lines are the pumps and the generated lines are labeled by the corresponding indices. The black arrows illustrate the degenerate (dashed) and non-degenerate (solid) four-wave mixing process. (c) The power conversion efficiency $\eta_{0\pm}$ scales quadratically with initial mechanical pump power in the waveguide $P'_{0\pm}$. The slope of the black line is the fitted $\Gamma$. The shaded regions are the convex hulls of the uncertainty on the measurement.}
\end{figure}

The FWM measurement setup is shown in Fig.~\ref{Fig:Nonlinear}a. Two strong microwave pump tones, driven at frequencies $f_{0\text{-}}$ and $f_{0\text{+}} = f_{0\text{-}} + \Delta f$, are combined in a power splitter and applied to the input IDT of a delay line. Nonlinear processes in the lithium niobate phononic waveguide lead to cascaded FWM and the generation of comb lines. Finally, the FWM signal is sent from the output IDT to a real-time spectrum analyzer (RSA, Rhode \& Schwarz FSW) for read-out.

We show a typical measured microwave power spectrum at the RSA in Fig.~\ref{Fig:Nonlinear}b where $\Delta f = \SI{10}{\mega\hertz}$. The waveguide length of this device is $L=\SI{800}{\micro\meter}$. The two central lines are at the pump frequencies and are centered around $\SI{3.23}{\giga\hertz}$. Both microwave source output powers are set to $6$~dBm. Due to the frequency dependence of the transducer efficiency $T\left(\omega\right)$, the two pump powers are not equal in the mechanical waveguide as well as at the RSA. We observe a comb of lines  spaced by $\Delta f$ on either side of the two pump tones which we label with index $n=\pm 1$, $\pm 2$, and so on. These are generated in the phononic waveguide from degenerate FWM as well as consecutive non-degenerate FWM between adjacent lines. The slowly varying amplitude $A_n$ of the mechanical wave for each comb line $n$ is governed by (Appendix~\ref{sec:FWM_theory}) 
\begin{equation}
    \partial_z A_n = -\frac\alpha 2 A_n +i\gammam \sum_{p+q-m = n}  A_p A_q A_m^*,
\end{equation}
where $\gammam$ is the modal nonlinear coefficient. $A_n(z)$ is normalized such that $P_n(z) = |A_n(z)|^2$ corresponds to the mechanical power. The spatial evolution of $A_n(z) $ is governed by the propagation loss $\alpha  $, and the FWM process between all components that satisfy the energy conservation. We have ignored phase-matching which is negligible for a waveguide length $L\sim \SI{1}{\milli\meter} $.

The nonlinear equation of evolution can be analytically solved in the low-power limit, where pump depletion and parametric amplification are negligible. A simple relationship between the power in adjacent lines holds (see Appendix~\ref{sec:FWM_theory}):
\begin{eqnarray}
    \eta_n &\equiv& \left|\frac{A_{n+1}}{A_n}\right|^2 =\left|\frac{A_{-n-1}}{A_{-n}}\right|^2 =  \frac {4\Gamma} {(n+1)^2} P_{0-}' P_{0+}',
\end{eqnarray}
for $n\ge 1$, where the $(n+1)$-th line is mainly generated by non-degenerate FWM of the $n$-th line and the two pumps. $P_{0-}'$ and $P_{0+}'$ are pump powers at the beginning of the waveguide. We have defined an effective FWM coefficient
\begin{equation}
\label{eq:Gamma-gamma}
    \Gamma = (\gammam L_\text{eff})^2,
\end{equation}
where $ L_\text{eff} = [1-\exp(-\alpha L )]/\alpha \approx \SI{0.53}{\milli\meter}$ is the effective length~\cite{agrawal2000nonlinear}. On the other hand, $ A_{\pm 1} $ are mainly created by degenerate FWM with corresponding efficiencies
\begin{equation}
\label{eq:eta0pm}
    \eta_{0\pm} \equiv \left|\frac{A_{\mp1}}{A_{0\pm}^*}\right|^2 = \Gamma P_{0\mp}'^2.
\end{equation}
In the experiment, microwave powers at different frequencies can be directly measured to calculate the corresponding mechanical powers in the waveguide, allowing us to further calculate $\Gamma$ and infer $\gammam$ from Eq.~\ref{eq:Gamma-gamma}. 

To extract the effective FWM coefficient $\Gamma $, we plot the measured $\eta_{0\pm} = P_{\mp1}/P_{0\pm}$ versus $P_{0\mp}'^2$ in Fig.~\ref{Fig:Nonlinear}c. By taking the mean of $\eta_{0\pm}/P_{0\mp}'^2$, we obtain $\Gamma=\SI{10 \pm 5}{ \per\milli\watt \squared}$ for the blue dataset and $\Gamma=\SI{20 \pm 10}{ \per \milli \watt\squared}$ for the red dataset. The large  uncertainty on $\Gamma$ mostly comes from the  uncertainty on the transmission $T$ of a single IDT, which is a function of frequency and cannot be directly measured. We have also ignored reflections $R$ at the output IDT. Only $\sim 30\%$ of the acoustic power gets reflected, $\sim 50\%$ of which has already decayed due to propagation loss. The contribution to $\Gamma$ from reflections is therefore negligible compared to the uncertainty. To eliminate a possible contribution from the nonlinearity of the measurement setup, we measure the effective FWM nonlinear coefficient $\Gamma_\text{s}$ of the measurement setup only, where the device is bypassed (Appendix~\ref{sec:nonlinear_setup}). The measured $\Gamma_\text{s}$ is more than four orders of magnitude smaller than $\Gamma$. This shows that the nonlinearity in our experiment is mechanical in nature.

Based on the measured $\Gamma$, we estimate the modal nonlinear coefficient $\gammam \sim  \SI{7}{\per\milli \watt\per\milli \meter}$, calculated from $\Gamma\sim \SI{15}{ \per \milli \watt\squared}$ and $L_\text{eff} = \SI{0.53}{\milli\meter}$. The modal nonlinear coefficient allows us to further estimate the pump power required for parametric amplification, where $ \eta \gtrsim 1$. Using Eq.~\ref{eq:eta0pm}, we obtain
\begin{equation}
     P' \gtrsim 1/\sqrt{\Gamma} = 1/(\gammam L_\text{eff}) \sim \SI{0.3}{\milli\watt}.
\end{equation}
By increasing the waveguide length, $L_\text{eff} $ can be increased to reduce the threshold power until $ L_\text{eff} \approx 1/\alpha $, where $ P' \sim \alpha/\gammam $. Cooling down the waveguide could reduce the propagation loss $\alpha$ and thus decrease the threshold power by at least one order of magnitude. We note that the relation between the FWM efficiency and the pump power is obtained in the low efficiency limit, which makes it invalid for $\eta \gtrsim 1$. By solving the coupled amplitude equations for degenerate FWM in a lossy waveguide, the same condition $P' \gtrsim 1/(\gammam L_\text{eff}) $ is obtained (Appendix~\ref{subsec:degenerate-FWM-gain}). Parametric amplification with mechanical FWM nonlinearity will be the subject of future exploration. 

\section{Conclusion and outlook}
\label{sec:conclusion}

We have introduced and demonstrated a platform for gigahertz-frequency phononic integrated circuits. LiSa's two essential properties -- index guiding due to higher speed of sound in sapphire, and the large piezoelectric coefficient of LN, enable the demonstrations in this work. In particular, we have shown compact interdigital transducers that can efficiently excite guided waves in a $\SI{1}{\micro\meter}$ wide waveguide at room temperature. To demonstrate some of the capabilities of this platform, we have realized delay lines, racetrack resonators, meander line waveguides, and studied nonlinear phononic effects that manifest in these structures. Measurements at cryogenic temperatures of resonators show  that $Q \sim 50,000$ and linewidths of $\sim\SI{70}{\kilo\hertz}$ are achievable at $\SI{4}{\kelvin}$.

Phononic integrated circuits on the LiSa platform provide an attractive route to implementing coherent microwave-frequency devices. At room temperature, the higher quality factors and large delays have long made acoustic delay lines attractive~\cite{lu2020ghz,lu2019gigahertz,vidal2017delay,campbell2012surface,coldren1976surface}. Our approach enables compact delay lines that allow phonons to propagate in a meander over a distance much longer than the extents of the chip. Our measurements show that the quality factors at room temperature are not limited by leakage of mechanical energy into the sapphire substrate. In addition to allowing for longer waveguides and easier integration, the unreleased nature of these circuits may greatly improve their robustness and power handling ability. Moreover, our platform also supports simultaneous guiding of low-loss optical waves (Appendix~\ref{sec:high-Q-optics}). Because both the acoustic and optical waves are highly co-localized in the waveguide, it is possible to obtain large optomechanical interaction rates~\cite{safavi2019controlling}. This opens up the possibility for realizing integrated Brillouin devices~\cite{eggleton2019brillouin,van2015interaction,liu2019electromechanical}, as well as acousto-optic modulation and beam steering~\cite{sarabalis2018optomechanical,li2019electromechanical,jiang2020efficient,shao2020integrated,sarabalis2020acousto} in an efficient and scalable way.

The demonstrated phononic platform may also find applications in emerging microwave quantum technologies. There have been numerous proposals for using phonons in quantum networks~\cite{habraken2012continuous,lemonde2018phonon,kuzyk2018scaling,fang2016optical,Schuetz2015universal,neuman2020phononic} to interconnect physically distant resonators or qubits, to realize time-delayed quantum feedback control~\cite{grimsmo2015time,pichler2016photonic,andersson2019non} and tensor network state generation~\cite{pichler2017universal}. The demonstrated gigahertz-frequency phononic circuits can be coupled to superconducting qubits via piezoelectric transducers~\cite{satzinger2018quantum,moores2018cavity,arrangoiz2018coupling,bienfait2019phonon,sletten2019resolving,arrangoiz2019resolving}. Long delays and high quality factors, and the compatibility of sapphire with high-$Q$ superconducting circuits make LiSa a promising platform for such hybrid quantum systems. 

To realize these advances, further work is needed to understand the sources of dissipation at cryogenic temperatures. Additionally, it is important to implement transducers that are more efficient by carefully engineering away loss channels associated with slab and bulk acoustic wave radiation. We expect that with such advances, the demonstrated platform will enable quantum and classical phononic circuits and systems with capabilities beyond the reach of current technologies.

\begin{acknowledgments}
The authors would like to thank Jason F. Herrmann and Hubert Stokowski for assistance on device fabrication, and Kevin Multani, Agnetta Y. Cleland and E. Alex Wollack for technical support. W.J. would like to thank Raphaël Van Laer for helpful discussions. This work is supported by the David and Lucile Packard Fellowship and by the U.S. government through the National Science Foundation (NSF) (1708734, 1808100), Airforce Office of Scientific Research (AFOSR) (MURI No. FA9550-17-1-0002 led by CUNY). Device fabrication was performed at the Stanford Nano Shared Facilities (SNSF) and the Stanford Nanofabrication Facility (SNF). SNSF is supported by the National Science Foundation under award ECCS-1542152.
\end{acknowledgments}

\appendix

\section{Device fabrication}
\label{sec:fab}

The fabrication process starts with $\sim \SI{500}{\nano\meter}$ thin-film LN on sapphire. The rib waveguide is patterned with electron beam lithography (EBL) using a negative resist (HSQ, FOx-16) and developed with $25\%$ TMAH. 
The LN is physically etched by $300$ nm using argon ion milling~\cite{jiang2019lithium}. For the slab removal, the patterning is done with positive resist (SPR3612) photolithography after which the LN is ion-milled. The samples are cleaned with $3:1$ piranha~\cite{hill2013nonlinear}.
We adopt two partial-etch steps instead of one full-etch step to avoid possible delamination between the LN mechanical waveguide and the sapphire substrate. The waveguide sidewall is tilted by approximately 11 degrees from the vertical direction, which is taken into account in the waveguide and IDT simulations.%

The IDT metallization layer is patterned by EBL liftoff using a positive resist (CSAR $6200.13$). A $\SI{10}{\nano\meter}$ aluminum layer is evaporated after the spin coat as a conductive layer for the EBL, which is then removed by 10:1 diluted hydrofluoric acid before the CSAR development. To make the metal climb the sidewalls from the sapphire to the LN, we evaporate $55$ nm aluminum on the sample at three different angles, perpendicular to the waveguide, and at $-66^\circ,66^\circ,0^\circ$ from the vertical direction. If only the two tilted evaporations are used, shadow from the first layer would prevent the electrical contact between the two layers. The chip is soaked one hour in $80^\circ $C NMP for liftoff. Finally, we use standard photolithography to pattern the bus electrodes and contact pads for the microwave probes, evaporate $150$ nm of aluminum, and liftoff in NMP.

\section{IDT admittance simulation}
\label{app:IDT_admittance}
To simulate the IDT admittance, we use the same model as shown in Fig.~\ref{Fig1:design}a with a geometry that is identical to the measured IDT in Fig.~\ref{Fig:IDT-measurement}. We find that a material loss that corresponds to an intrinsic quality factor of $Q_\text{i} = 400$ and scaling of the piezoelectric coefficient by $67\% $ are required to obtain better matching between simulation and measurement, similar to previously reported values for thin-film LN transducers~\cite{sarabalis2020s,dahmani2020piezoelectric}.

\begin{figure}[htbp]
  \includegraphics[scale=1]{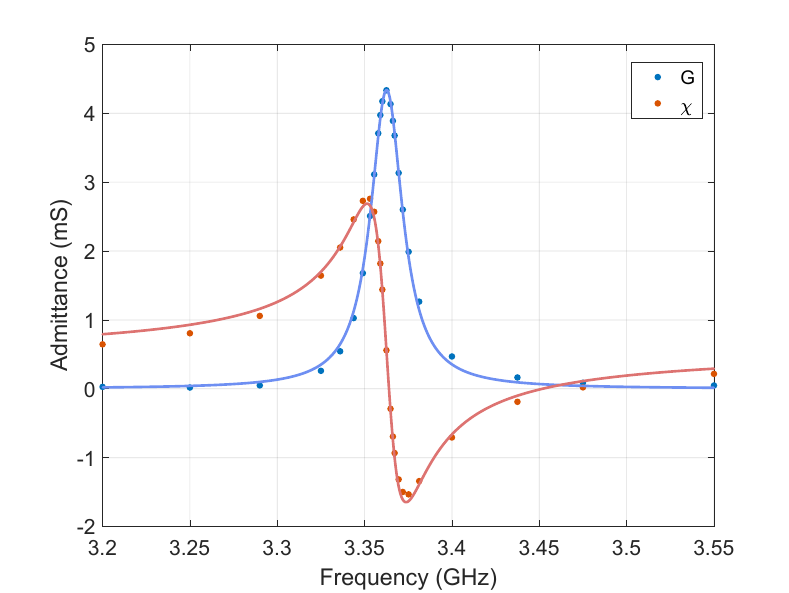} 
\caption{\label{Fig:IDT_sim} Simulated IDT admittance. Solid lines are fits to the simulated data. }
\end{figure}

The simulated conductance $G$ and susceptance $\chi$ are shown in Fig.~\ref{Fig:IDT_sim} as blue and red dots. By fitting the simulated conductance, we extract the peak conductance $G_0 = 4.3~\text{mS}$ and the IDT bandwidth $\gamma/2\pi = 22.3~\text{MHz}$, which quantitatively agree with measurements.

\section{Deembedding IDT scattering parameters and waveguide propagation loss}
\label{sec:deembed}

\begin{figure}[ht]
  \includegraphics[scale=1]{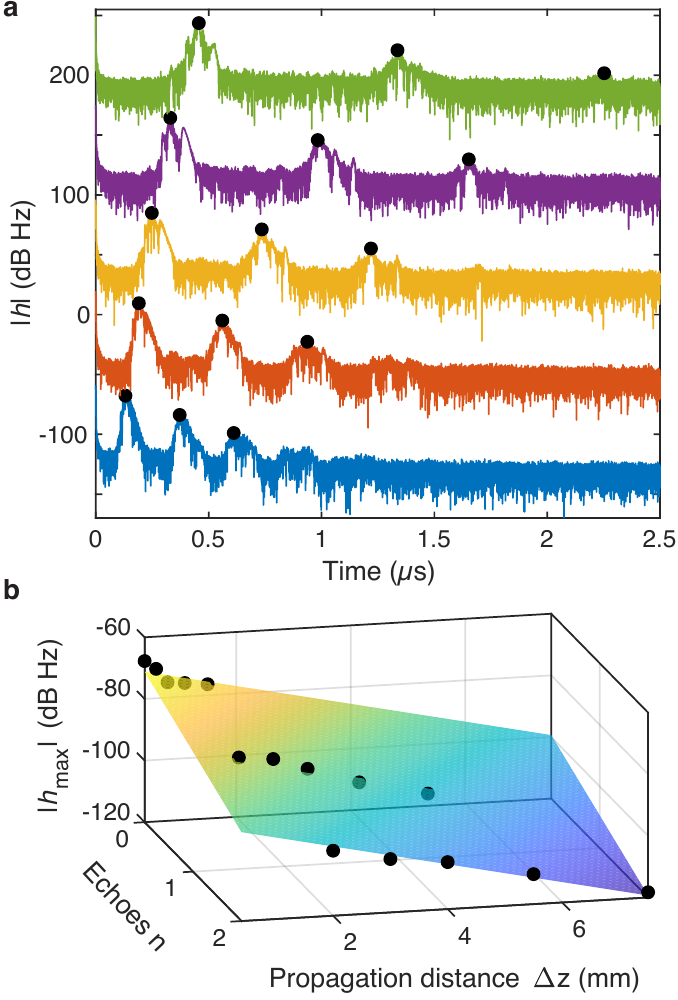} 
\caption{\label{Fig:alpha_plane_fit} (a) Time domain impulse response $\left| h\right|$ for $5$ delay lines with waveguide lengths $L=0.4$, $0.6$, $0.8$, $1.1$ and $1.5$ mm (from bottom to top). The black dots are at the maximum of the peaks representing the single transit and first two echoes. The responses are manually shifted by $\SI{80}{\deci\bel\hertz}$ for clarity. (b) The maximum impulse response is fit with a plane against number of echoes and propagation distance.}
\end{figure}

We would like to extract IDT scattering parameters (reflection and conversion efficiency) as well as propagation losses in the waveguide from $S_{21}$ measurements using a technique similar to the one presented in~\cite{sarabalis2020s,dahmani2020piezoelectric}. We analyze the impulse response $h_{21}(\tau)$ of the delay line (Fig. \ref{Fig:alpha_plane_fit}a), found by taking the discrete Fourier transform of $S_{21}(\omega)$. In the impulse response, we observe a series of peaks separated by a delay $\Delta \tau = 2L/v_g$, where $L$ is the length of the waveguide and $v_g$ is the group velocity of the mechanical wave. While the first peak corresponds to a single transit of the mechanical wave through the waveguide, the subsequent peaks are due to echoes from reflections at the transducers back into the waveguide. These therefore have later arrival times. If the peaks are well resolved (that is, the delay line is long enough), we can write for each peak:
\begin{equation}
    \left|h(n)\right| = t_{\text{m}\mu}^2 r^{2n} e^{-\frac{\alpha}{2} \Delta z_n}.
\end{equation}
Here, $\alpha$ is the propagation loss, $t_{\text{m}\mu}$ is the scattering parameter between the microwave input signal and the desired propagating mode of the mechanical waveguide, $r$ is the mechanical amplitude reflection coefficient at the transducer-waveguide interface, $n$ is the number of echoes and $\Delta z_n = (2n + 1) L $ is the propagation distance. When considering the power instead of the amplitude impulse response of the device, this expression becomes:
\begin{equation}
\label{eq:h_deembed}
    \left|h(n)\right|^2 = T^2 R^{2n} e^{-\alpha \Delta z_n}.
\end{equation}
Here, $T=|t_{\text{m}\mu}|^2$ is the conversion efficiency of the IDT and $R=|r|^2$ is the mechanical power reflection coefficient. To deembed $R$ and $\alpha$ from $\left| h\right|^2$, we measure $5$ devices with waveguide lengths $L=0.4$, $0.6$, $0.8$, $1.1$, $1.5$ mm and take the maximal value $\left| h_\text{max}\right|^2$ of the power impulse response for three peaks each (see Fig. \ref{Fig:alpha_plane_fit}a). We express all quantities in Eq.~\ref{eq:h_deembed} in logarithmic scale  and fit a plane to:
\begin{equation}
    2\ln \left| h_\text{max}\right| = 2\ln T + 2n\ln R - \alpha \Delta z_n,
\end{equation}
where $\ln T$, $\ln R$ and $\alpha$ are the free parameters. A rendering of this plane fit is shown in figure \ref{Fig:alpha_plane_fit}b. We find $\alpha = 4.0 \pm \SI{0.2}{\deci\bel\per\milli\meter}$ and $R = 32 \pm 8$\%.

The propagation loss $\alpha$ found here matches the one found for racetrack resonators in section \ref{sec:racetrack}.
Finally, we can use this propagation loss to infer the average conversion efficiency $T$ for an individual transducer pair. By filtering the $S_{21}$ measurement of a delay line to keep only the single-transit response ($n=0$), we obtain $T$ from:
\begin{equation}
    2\ln T = 2\ln \left| S_\text{max}\right| + \alpha L.
\end{equation}

\section{Filtering the scattering parameters of the acoustic racetrack resonator}
\label{app:filter}
When measuring the scattering parameters of the acoustic racetrack resonator, the signal from the mechanical resonances of the racetrack interferes with the microwave crosstalk and the mechanical reflections at the IDTs. The former typically leads to a broad background as depicted in Figure~\ref{Fig:IDT-measurement}b, while the latter results in fringes which can be seen in Figure~\ref{Fig:IDT-measurement}c. To reduce the effect of this interference, the scattering parameters are inverse Fourier transformed, filtered in time-domain, and transformed back. The filtering involves manually setting the time-domain impulse response to zero for a short time interval.%

In the through configuration, we filter out the impulse response for small delays $<\SI{50}{\nano\second}$ which are due to microwave crosstalk and do not involve contributions from the slower mechanical waves. Similarly, we also filter out the impulse response for delays $\SI{140}{\nano\second}<\tau<\SI{330}{\nano\second}$ to remove echoes from mechanical reflections between the input and output IDTs. We do not do any filtering for larger delays where the impulse response has contributions from mechanical waves propagating through the racetrack resonator. In the drop configuration, we filter out the impulse response for delays $<\SI{190}{\nano\second}$, before any of the guided mechanical waves have had time to reach the output IDT.

\section{Waveguide bending loss simulation}
\label{sec:bending}

In this section, we show that the quality factor of our acoustic racetrack resonator is far from being limited by the simulated bending loss.

Due to the anisotropic elastic and piezoelectric properties of lithium niobate, the bending loss is position-dependent along the arc. To avoid simulating the full ring or racetrack structure, we take a thin slice of sector on the arc of which the corresponding central angle $\theta \ll 1$. A mechanical perfectly-matched layer is added to the sapphire substrate. A cyclic symmetry boundary condition is applied to the two end surfaces of the sector,
\begin{equation}
    \bm u_\text d = \bm R \cdot \bm u_\text s e^{-i m \theta}.
\end{equation}
$ \bm u_\text s $ and $ \bm u_\text d $ are the displacement field on the source and destination surfaces, $\bm R$ is the rotation matrix from the source surface to the destination surface, and $m = k R$ is the azimuthal mode number for a resonating mode with wavevector $k$ in a ring with radius $R$. Note that this approximates the crystal as if it also had cyclic symmetry, under which the material properties are suddenly rotated by $\theta$ when the wave propagates from one sector to the next sector.

\begin{figure}[ht]
  \includegraphics[scale=1]{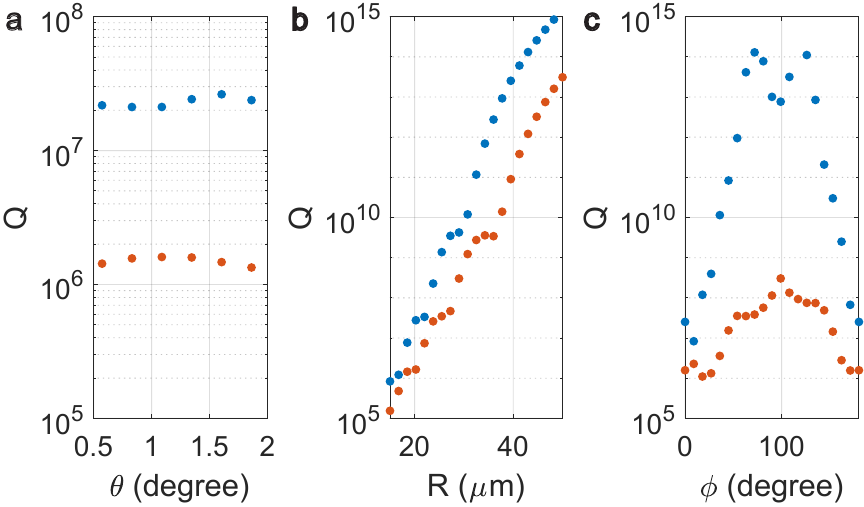} 
\caption{\label{Fig:bending_loss} Simulated bending-loss limited quality factor $Q$ for the first (blue, quasi-Love) and second (red, quasi-Rayleigh) guided mechanical mode as a function of (a) central angle $\theta$ of the simulated sector, (b) bending radius $R$, and (c) crystal orientation $\phi$. }
\end{figure}

To show that the simulated bending loss is independent of the artificially chosen $\theta$, we fix the bending radius of the arc to be $R = \SI{20}{\micro\meter}$, the orientation of the crystal such that the crystal $ Y $ axis is parallel to the direction of propagation, and change the central angle $\theta$ of the sector. The simulated quality factor is plotted in Fig.~\ref{Fig:bending_loss}a. We observe minor changes in the simulated $Q$, likely due to variations in the mesh of the model. $\theta$ is kept to be smaller than $1.5$ degree for all subsequent simulations. When we sweep the values of the bending radius $R$, an exponential dependency of bending loss versus $R$ is observed in Fig.~\ref{Fig:bending_loss}b, similar to Ref.~\cite{fu2019phononic}.

The crystal orientation dependency of the bending loss is shown in Fig.~\ref{Fig:bending_loss}c. $\phi$ is defined as the angle between the crystal $Y$ axis and the direction of propagation. The bending radius is fixed at $R =  \SI{20}{\micro\meter}$. We observe a huge and nontrivial variation of simulated $Q$, showing that the bending loss changes drastically along the arc. The actual bending loss limited quality factor can be calculated from the total bending loss per round trip, which can be integrated from the simulated orientation-dependent bending loss,
\begin{equation}
\label{eq:Q-ring}
    Q_\text{bending} = \frac{ \int_0^{2\pi} d\phi / v(\phi) }{ \int_0^{2\pi} d\phi / [Q(\phi)v(\phi)]},
\end{equation}
where $ v(\phi) $ is the orientation-dependent group velocity. Eq.~\ref{eq:Q-ring} can be generalized for racetracks. As a result, the actual $Q_\text{bending}$ lies between the minimal and maximal simulated $Q(\phi)$.  All the fabricated acoustic racetrack resonators have bending radii above $\SI{20}{\micro\meter} $. The measured quality factors at both room temperature and $\SI{4}{\kelvin}$ are thus far from being limited by bending loss.

\section{Cascaded four-wave mixing in a lossy waveguide}
\label{sec:FWM_theory}

In this section, we derive the power ratio between adjacent frequencies from cascaded four-wave mixing in a phase-matched lossy waveguide.

We follow the derivation in Ref.~\cite{ng2005cascaded,agrawal2000nonlinear} to obtain the set of equations governing the evolution of amplitudes at different frequencies. The slowly varying envelope $A$ of a propagating wave with central frequency $\omega_0$ can be expanded with envelopes $A_n $ at different frequencies $\omega_n$,
\begin{eqnarray}
    A(z,t)e^{i(k_0 z - \omega_0 t)} &=& \sum_n A_n(z) e^{i( k_n z - \omega_n t )},\\
    k_n &=& \beta(\omega_n),\\
    A &=& \sum_n A_n e^{i(k_n' z - \omega_n' t)},
\end{eqnarray}
where $ k_n' = k_n - k_0 $ and $ \omega_n' = \omega_n - \omega_0 = n\Delta\omega$. $ \Delta \omega$ is the frequency spacing of the four-wave mixing process and $n$ takes integer values.

The equation that governs the amplitudes $ A_n $ is nearly identical to the optical four-wave mixing case and reads
\begin{equation}
    \sum_n \partial_z A_n e^{i(k_n' z - \omega_n' t)} = -\frac \alpha 2 A + i\gammam |A|^2 A.
\end{equation}
We have added a damping term to account for the propagation loss $\alpha$. The amplitude $A$ is normalized such that $|A|^2$ is the power. $\gammam$ is the modal nonlinear coefficient and has a dimension of (power$\cdot$length)$^{-1}$. By further substituting $A$ with the sum of all frequency components and collecting terms with the same frequency, we obtain
\begin{equation}
    \label{eq:FWM-An-evolution}
    \partial_z A_n = -\frac\alpha 2 A_n +i\gammam \sum_{p+q-m = n}  A_p A_q A_m^* e^{ i\Delta k z},
\end{equation}
where $\Delta k \equiv k_p' + k_q' - k_m' - k_n'$ accounts for the phase mismatch. We see that the spatial evolution of the envelope $A_n$ is damped by the propagation loss $\alpha$, and is also generated from four-wave mixing of all frequency components satisfying energy conservation.

The value of the mismatch $\Delta k$ depends on the dispersion of the mechanical waveguide. We perform a polynomial fit to the simulated dispersion relation $\beta(\omega)$ over frequencies $2\sim\SI{3}{\giga\hertz} $ and extract the dispersion coefficients
\begin{equation}
    \beta_k\equiv d^k\beta/d\omega^k |_{\omega=\omega_0}
\end{equation}
at $\omega_0/2\pi = 3.4$ GHz. We obtain $\beta_2 = 1.1 ~\SI{}{\nano\second\squared\per\radian\per\milli\meter}$. Expressing $\Delta k = \frac12 \beta_2 \Delta\omega^2 (p^2+q^2-m^2-n^2) + O(\Delta\omega^3)$ we find that for a four-wave mixing bandwidth of $n\Delta\omega/2\pi \sim \SI{100}{\mega\hertz}$,
\begin{equation}
    \Delta k \sim \frac{1}{2}\beta_2 n^2\Delta\omega^2 \sim 2\pi\times \SI{34}{\per\meter}.
\end{equation}
The phase-matching condition becomes important for a waveguide of length $L\gtrsim \SI{30}{\milli\meter} $, which is far longer than the measured waveguides with $L\lesssim \SI{1}{\milli\meter}$. We conclude that the effects of dispersion can be safely ignored in these waveguides and take $\Delta k = 0$ from now on.

We move on to consider the measurement scheme used in Sec.~\ref{sec:FWM}, where two strong pump tones are launched in the mechanical waveguide at frequencies $\omega_{0-}$ and $\omega_{0+} = \omega_{0-} + \Delta\omega$. Since the pump amplitudes $ A_{0-} $ and $A_{0+} $ are much larger than any other $A_n$, most terms in the sum in Eq.~\ref{eq:FWM-An-evolution} can be ignored except terms that involve two amplitudes from $A_{0-}$ and $A_{0+}$ (the validity of this approximation is evaluated in a self-consistent way). In addition, we assume that the FWM drops off quickly so that the contributions to $A_{\pm n}$ from $A_{\pm(n+1)}$ are negligible. We also ignore self- and cross-phase modulation for now, which would modify the phase-matching condition. As a result,
\begin{eqnarray}
    \partial_z A_{n} &=& -\frac\alpha 2 A_{n} + 2 i\gammam A_{0+} A_{0-}^* A_{n-1},\label{eqn:comb-amp-pos-n}\\
    \partial_z A_{-n} &=& -\frac\alpha 2 A_{-n} + 2 i\gammam A_{0-}  A_{0+}^* A_{-n+1},\label{eqn:comb-amp-neg-n}
\end{eqnarray}
where $n$ is a positive integer. $A_{\pm 1}$ need to be considered separately, since they are generated solely from $A_{0-}$ and $A_{0+}$,
\begin{eqnarray}
    \partial_z A_{1} &=& -\frac\alpha 2 A_{1} + i\gammam A_{0+}^2 A_{0-}^*,\\
    \partial_z A_{-1} &=& -\frac\alpha 2 A_{-1} +  i\gammam A_{0-}^2  A_{0+}^*.
\end{eqnarray}
This set of equations shows that the FWM lines are generated in a cascaded fashion from the central two pump lines $A_{0\pm}$.

Given that the measured power in the FWM generated amplitudes are much smaller than the pump, we ignore pump depletion, and the amplitudes of the pump fields are $ A_{0-}(z) = A_{0-}' \exp(-\alpha z/2) $, $ A_{0+}(z) = A_{0+}' \exp(-\alpha z/2) $, where $ A_{0-}' $ and $A_{0+}'$ are the initial amplitudes at the beginning of the waveguide $z = 0 $. We expect all amplitudes at different frequencies to share the same exponential decay factor as there should not be a strongly frequency-dependent loss mechanism in the waveguide. This can be taken into account by the following substitution
\begin{equation}
    A_n = A_n' e^{-\alpha z/2},
\end{equation}
after which the evolution of $A_n'$ is now governed by
\begin{eqnarray}
    \partial_z A_{n}' &=&  ig A'_{n-1} e^{-\alpha z},\label{eqn:comb-amp-pos-n-lossy}\\
    \partial_z A_{-n}' &=& ig^* A'_{-n+1} e^{-\alpha z}.\label{eqn:comb-amp-neg-n-lossy}\\
    \partial_z A'_{\pm1} &=& i\gammam A_{0\mp}^{'*} A_{0\pm}^{'2} e^{-\alpha z}.\label{eqn:comb-amp-pm1}
\end{eqnarray}
We have defined $ g\equiv 2 \gammam  A_{0-}^{'*} A'_{0+}  $ for short. Under the initial condition $ A_n'(0) = 0$ for $n\neq 0$, the solutions are
\begin{eqnarray}
    A'_{n}(z) &=& A'_{n}(\infty) (1-e^{-\alpha z})^{n},
\end{eqnarray}
where
\begin{eqnarray}
    A'_{\pm1}(\infty)  &=& i\gammam A_{0\mp}^{'*} A_{0\pm}^{'2} /\alpha,\\
    A'_{-n}(\infty) &=& \left(\frac{ig^*}{\alpha} \right)^{n-1} \frac 1 {n!} A'_{-1}(\infty),  \\
    A'_{n}(\infty) &=& \left(\frac{ig}{\alpha} \right)^{n-1} \frac 1 {n!} A'_{1}(\infty).
\end{eqnarray}
We note that the cascaded FWM process is symmetric between lower and higher frequencies. %

The solutions allow us to calculate the power conversion efficiency from the $n$-th to the $(n+1)$-th frequency components,
\begin{eqnarray}
    \eta_n &\equiv& \left|\frac{A_{n+1}}{A_n}\right|^2 =\left|\frac{A_{-n-1}}{A_{-n}}\right|^2 =  \frac {4\Gamma} {(n+1)^2} P_{0-}' P_{0+}',
\end{eqnarray}
where we have defined an effective FWM coefficient
\begin{equation}
\label{eq:Gamma}
    \Gamma = \left( \frac{\gammam}{\alpha} \right)^2 (1-e^{-\alpha z})^2, %
\end{equation}
which is a function of the propagation length $z$ and has a dimension of (power)$^{-2}$. $P_{0-}' = |A_{0-}'|^2 $ and $P_{0+}'=|A_{0+}'|^2$ are the pump powers launched into the waveguide. We observe that the power ratio between adjacent FWM components is proportional to the coefficient $\Gamma$ and the two pump powers. 

$A_{\pm 1}$ are generated from a different process, and it is more suitable to define $\eta_{0\pm}\equiv |A_{\mp1}/A_{0\pm}^*|^2$, leading to
\begin{eqnarray}
\label{eqn:t1-expr}
    \eta_{0\pm} & = & \Gamma P_{0\mp}'^2.
\end{eqnarray}

There are clearly two regimes for the effective FWM coefficient $\Gamma$. For a short propagation length $z = L \ll 1/\alpha$,
\begin{equation}
    \Gamma = \left( \gammam L \right)^2.
\end{equation}
The conversion efficiency grows quadratically with respect to $L$. In the other limit where $ L\gg 1/\alpha $,
\begin{equation}
    \Gamma = \left( \frac{\gammam}{\alpha} \right)^2.
\end{equation}
The efficiency grows to a fixed value, while all components decay exponentially along the waveguide due to the propagation loss. In the general case, we could define an effective length $ L_\text{eff} = [1-\exp(-\alpha z )]/\alpha $, and thus
\begin{equation}
    \Gamma = (\gammam L_\text{eff})^2.
\end{equation}

The outgoing microwave powers of different frequency components $ P_{\mu,n} = |t_{\text m \mu}(\omega_n)|^2 P_n = T(\omega_n) P_n $ can be directly measured, allowing us to estimate the mechanical powers at the beginning of the waveguide $ P_{0\pm}'$, at the end of the waveguide $P_n$, and extract $\Gamma $. More specifically,
\begin{equation}
    \eta_{0\pm} = \frac{P_{\mp1}}{P_{0\pm}} = \frac{P_{\mu,\mp1}}{P_{\mu,0\pm}} \frac{ T(\omega_{0\pm}) }{T(\omega_{\mp1})},
\end{equation}
where $ T(\omega) $ takes into account the frequency dependency of the IDT conversion efficiency. As a result,
\begin{equation}
\label{eq:Gamma_calibration}
    \Gamma = \frac{\eta_{0\pm}}{P'^2_{0\mp}} = \frac{P_{\mu,\mp1}}{P_{\mu,0\pm}} \frac{ T(\omega_{0\pm}) }{T(\omega_{\mp1})} \left( \frac{\eta_\text{m} T(\omega_{0\mp}) e^{-\alpha L}}{P_{\mu,0\mp}} \right)^2.
\end{equation}
$\eta_\text{m} \approx 60$ \% is the output insertion loss between the output IDT and the RSA.

We end this section with a quick estimation of the self phase modulation (SPM) and cross phase modulation (XPM). The measured nonlinear coefficients are $\Gamma \sim \SI{15}{\milli\watt\per\milli\watt\cubed} $ and $\gammam \sim \SI{7}{\per\milli\watt\per\milli\meter}$. For a pump mechanical power $P \sim \SI{10}{\micro\watt}$ in a waveguide of length $ L\sim 1/\alpha \sim \SI{1}{\milli\meter} $, the accumulated phase for SPM and XPM is
\begin{equation}
    \phi \sim \gammam P L \sim 0.07 \ll \pi.
\end{equation}
As a result, we conclude that with the measured FWM nonlinearity and the pump power used, the SPM and XPM effects are negligible.

\section{Degenerate parametric gain in a lossy waveguide}
\label{subsec:degenerate-FWM-gain}

In the previous section, we discussed the approximated solutions for cascaded FWM in the low efficiency limit. To observe parametric amplification, an efficiency $\eta \sim 1$ is anticipated, and the approximations are no longer valid. Here, we focus on degenerate FWM with loss, and derive the condition for obtaining significant parametric gain.

We consider the degenerate FWM process with pump field $ A_0 $, signal $A_1$, and idler $A_{-1}$. Using the same notations as defined in Appendix~\ref{sec:FWM_theory}, we start with a set of coupled amplitude equations for $A'_{\pm 1} $,
\begin{eqnarray}
    \partial_z A'_1 &= & i\gammam A_0^{'2} A_{-1}^{'*} e^{-\alpha z},\\
    \partial_z A'_{-1} &= & i\gammam A_0^{'2} A_{1}^{'*} e^{-\alpha z},
\end{eqnarray}
which are similar to Eq.~\ref{eqn:comb-amp-pm1}. Phase mismatch from group velocity dispersion, as well as SPM and XPM, are ignored. After eliminating $A'_{-1}$, we obtain a second-order differential equation for $A'_1$,
\begin{equation}
\label{eq:A1-lossy-nonlinear}
\partial_z^2 A'_1 + \alpha \partial_z A'_1 - (\gammam P'_0)^2 A'_1 e^{-2\alpha z} = 0.
\end{equation}
By assuming $ A'_1 \propto \exp[ a \exp(b z)] $ where $a$ and $b$ are constants to be determined, and substituting in Eq.~\ref{eq:A1-lossy-nonlinear}, we obtain $ b=-\alpha $, $ a = \pm \gammam P'_0/\alpha$, and
\begin{eqnarray}
    A'_{\pm1} &=& a_{\pm1} \exp \left[ - \frac{\gammam P'_0}{\alpha} \left(e^{-\alpha z}-1\right) \right]\nonumber\\
    &&+ b_{\pm1} \exp \left[ \frac{\gammam P'_0}{\alpha} \left(e^{-\alpha z}-1\right) \right],\nonumber\\
    &=&a_{\pm1} \exp \left(\gammam P'_0 L_\text{eff}\right)+ b_{\pm1} \exp \left( - \gammam P'_0 L_\text{eff}\right),
\end{eqnarray}
where $a_{\pm1}$ and $b_{\pm1}$ are constants determined from the boundary conditions. $ L_\text{eff} = [1-\exp(-\alpha z )]/\alpha $ is the effective length. Note that in the lossless limit,
\begin{equation}
    A'_{\pm1}|_{z\ll 1/ \alpha} = a_{\pm1} \exp \left(\gammam P'_0 z\right)+ b_{\pm1} \exp \left( - \gammam P'_0 z\right),
\end{equation}
which indicates a parametric gain $g = \gammam P'_0$ that agrees with the well-known result when the effective phase mismatch is ignored.

For a waveguide length $ L\gg 1/\alpha$, $L_\text{eff}\rightarrow 1/\alpha$,
\begin{equation}
    A'_{\pm1}(L) = a_{\pm1} \exp \left( \frac{\gammam P'_0}{\alpha} \right) + b_{\pm1} \exp \left( -\frac{\gammam P'_0}{\alpha} \right),
\end{equation}
showing that the final amplitudes $ A'_{\pm 1} $ converge to constant values, and $A_{\pm 1} $ simply decay with propagation loss $\alpha$.

Note that the initial amplitudes are $ A'_{\pm1}(0) = a_{\pm1}  +  b_{\pm1}$ and it is thus reasonable to identify $G = \exp \left( {\gammam P'_0} L_\text{eff} \right) $ as the amplification. Significant amplification is achieved when $ P'_0 \gtrsim 1/(\gammam L_\text{eff})$. Interestingly, using $\eta = \Gamma P^{'2}$ obtained in the low efficiency limit $ \eta \ll 1 $, the condition $ \eta \gtrsim 1 $ gives the same condition on pump power
\begin{equation}
    P' \gtrsim \frac{1}{\sqrt{\Gamma}} = \frac{1}{\gammam L_\text{eff}}.
\end{equation}

\section{Nonlinearity of the measurement setup}
\label{sec:nonlinear_setup}

To demonstrate that the measured FWM is from the device as opposed to the measurement setup, we bypass the device and carry out the same nonlinear measurements. The FWM nonlinearity of the setup is measured to be more than four orders of magnitude smaller than of the mechanical waveguide, making it negligible in the actual measurement.

\begin{figure}[ht]
\vspace{0.2in}
  \includegraphics[scale=1.05]{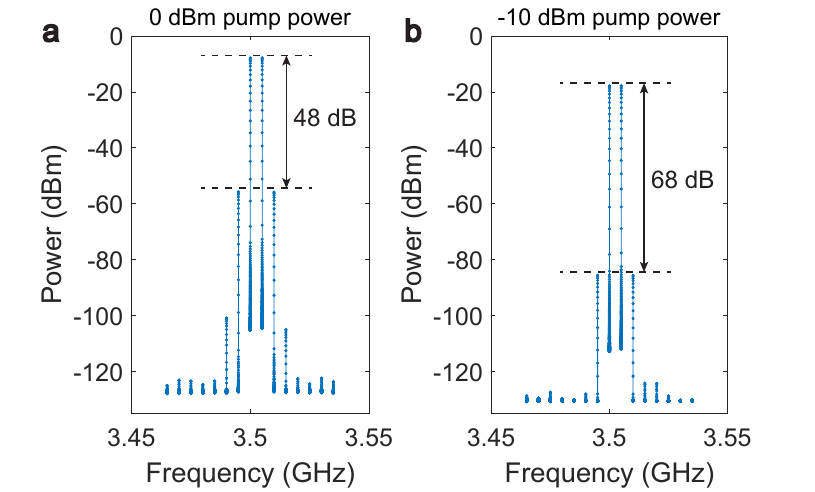} 
\caption{\label{Fig:setup-nonlinearity} Four-wave-mixing nonlinearity of the measurement setup, measured using two pump signals with (a) $0$ dBm and (b) $-10$ dBm output power from the signal generator.}
\end{figure}

Fig.~\ref{Fig:setup-nonlinearity} shows the measured spectrum at the RSA with two pump tones at $f_{0-} = \SI{3.5}{\giga\hertz}$ and $ f_{0+} = f_{0-} + \Delta f $. At pump power $ P_{0\pm} = 0 $ dBm, the first generated FWM line is $\SI{48}{\deci\bel}$ lower than the pump. For lower pump power at $-10$ dBm, the ratio becomes $\SI{68}{\deci\bel}$. This is expected from a FWM process where, for example, $ P_1 = \Gamma_\text{s} P_{0+}^2 P_{0-} $ hence $ P_1/P_{0-} = \Gamma_\text{s} P_{0+}^2 $. We define an effective FWM efficiency $\Gamma_\text{s}$ of the measurement setup, which can be directly compared to the measured $\Gamma$ as defined in Appendix~\ref{sec:FWM_theory} when the device is included. Using the measured powers at the RSA, we extract $\Gamma_\text{s} = \SI{550}{\per\watt\squared} = \SI{0.55}{\micro\watt\per\milli\watt\cubed}$ for both $0$ dBm and $-10$ dBm pump powers. We notice the presence of an insertion loss of $ \SI{-7.75}{\deci\bel} $. Taking into account the insertion loss, the actual $\Gamma_\text{s}$ lies between $ \SI{16}{\per\watt\squared} $ and $\SI{550}{\per\watt\squared} $ depending on where in the setup the FWM is happening. We further sweep $\Delta f$ from $\SI{1}{\mega\hertz}$ to $ \SI{50}{\mega\hertz} $. Relative variation of $\Gamma_\text{s}$ over different $\Delta f$ is measured to be less than $ 5\%$, indicating that the nonlinearity is relatively broadband. Noting that $ \Gamma_\text{s} \ll \Gamma \sim \SI{15}{\milli\watt\per\milli\watt\cubed} $, we thus conclude that the measured $\Gamma $ is dominantly from the IDT-waveguide-IDT device.

\section{High-Q optical ring cavities}
\label{sec:high-Q-optics}

The LiSa material platform is compatible with lithium niobate nanophotonics. Optical ring cavities with $\SI{70}{\micro\meter}$ radius were fabricated on the same chips as the phononic components and with waveguide dimensions identical to the mechanical waveguide. They were measured before the IDT metallization as well as at the end of the fabrication process.

\begin{figure}[ht]
\vspace{0.2in}
  \includegraphics[scale=0.9]{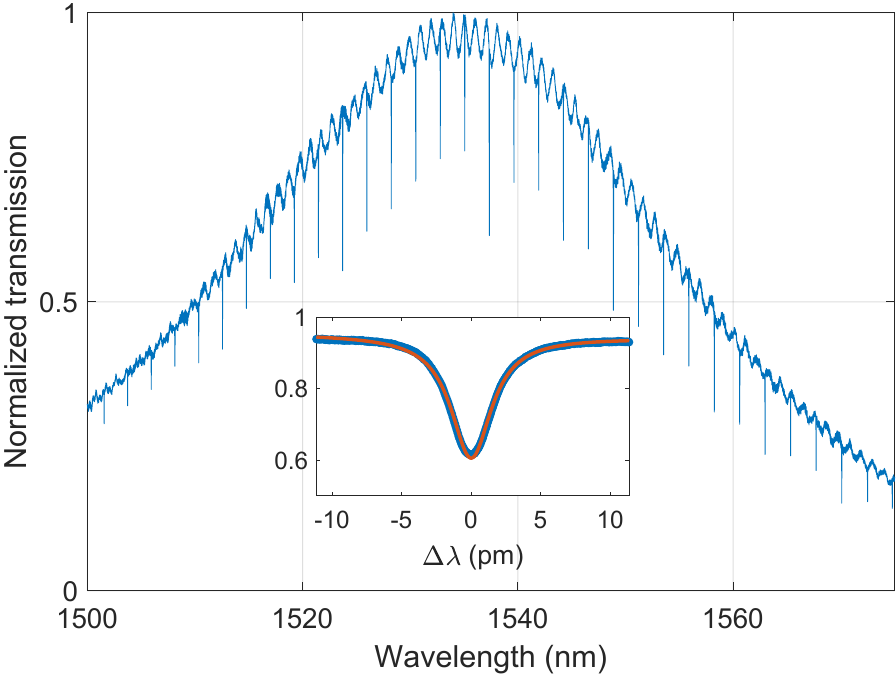} 
\caption{\label{Fig:optics} Transmission spectrum of an optical ring cavity fabricated on the same chip after the full fabrication process. Inset shows a typical mode from the ring cavity and the corresponding Lorentzian fit, giving a loaded quality factor $ Q = 4.3 \times 10^{5} $ and an intrinsic quality factor $ Q_\text i = 5.4\times 10^5 $.}
\end{figure}

A typical measured spectrum from a ring cavity is shown in Fig.~\ref{Fig:optics}. We extract an average loaded quality factor of $Q\sim 800,000$ before metallization and $Q \sim 500,000 $ at the end of the fabrication. We note that cladding the LN waveguide with silicon oxide is commonly adopted for reducing scattering loss from the LN surface, where quality factor above one million can be achieved on the LiSa platform~\cite{mckenna2019low,witmer2019chip}. Our optical waveguides and ring cavities are not cladded with oxide. The lower quality factors at the end of the fabrication are likely due to contamination from the metal liftoff resist, and could be potentially improved by adopting extra descum steps before the metal evaporation and after the liftoff.

\section{Bulk acoustic resonances}
\label{app:BAW}

\begin{figure}[ht]
\vspace{0.2in}
  \includegraphics[scale=1]{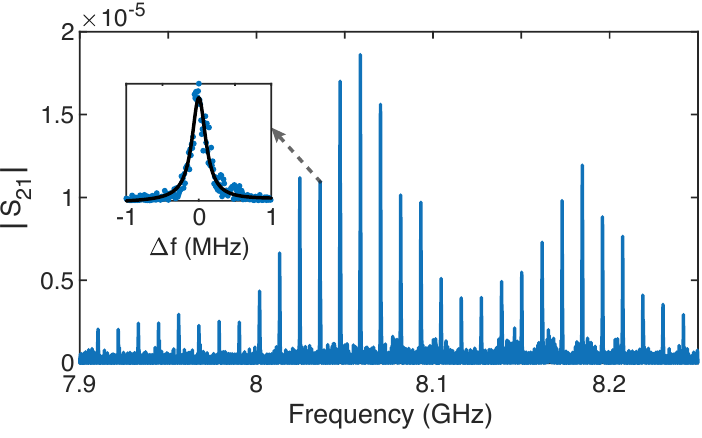} 
\caption{\label{Fig:BAW} $\left|S_{21}\right|$ at temperature $T=\SI{4}{\kelvin}$ of an IDT-waveguide-IDT device revealing bulk acoustic wave resonances. The inset shows a typical resonance with quality factor $Q=34,000$.}
\end{figure}

When characterizing the IDT-waveguide-IDT device at $\SI{4}{\kelvin}$, we observe evenly spaced modes around $ \SI{8}{\giga\hertz} $ in the transmission spectrum. Based on the free spectral range of $\SI{11.4}{\mega\hertz}$, we identify these modes as formed by the longitudinal bulk acoustic waves (BAW) that are mostly in the sapphire substrate, where the longitudinal wave speed of $ \SI{10.7}{\kilo\meter\per\second}$ and the substrate thickness of $ \sim \SI{500}{\micro\meter}$ predict an FSR of $ \sim \SI{10.7}{\mega\hertz}$. The longitudinal BAW speed along crystal $X$ in LN is $\SI{6570}{\meter\per\second}$. The corresponding wavelength is $\sim \SI{0.8}{\micro\meter}$ at $\sim \SI{8}{\giga\hertz}$. Maximal piezoelectric coupling between the BAW and the electric field occurs when half of the wavelength matches the LN thickness~\cite{han2016multimode}, which falls between the LN slab thickness $ t_\text s \approx \SI{200}{\nano\meter} $ and the total LN thickness $ \sim \SI{500}{\nano\meter} $. We would like to point out that the transducer is not specifically designed to transduce BAW, and the chip is not cleaned and mounted in a way that minimizes the scattering loss of BAW at the bottom surface of the chip.

BAWs in sapphire can be utilized for optomechanical applications~\cite{renninger2018bulk}. Moreover, the LiSa material platform is naturally compatible with superconducting circuits and could enable hybrid quantum systems with BAWs~\cite{chu2017quantum,chu2018creation} for hardware efficient quantum memory~\cite{pechal2018superconducting,hann2019hardware}.

\end{document}